\documentclass{article}

\usepackage[nonatbib,preprint]{neurips_2022}

\usepackage[utf8]{inputenc} 
\usepackage[T1]{fontenc}    
\usepackage{hyperref}       
\usepackage{url}            
\usepackage{booktabs}       
\usepackage{amsfonts}       
\usepackage{nicefrac}       
\usepackage{microtype}      
\usepackage{xcolor}         
\usepackage{tabto}    
\usepackage[superscript,biblabel]{cite}
\usepackage{setspace}  
\usepackage{titlesec}

\usepackage{subfig}
\usepackage{wrapfig}
\usepackage{tikz}
\usepackage{tikz-cd}
\usetikzlibrary{cd}
\usepackage{microtype}
\usepackage{graphicx}
\usepackage{booktabs} 
\usepackage{pifont}       
\usepackage{bbding}       %
\usepackage{fontawesome}  
\usepackage{jabbrv}

\usepackage{amsmath}
\usepackage{amssymb}
\usepackage{mathtools}
\usepackage{amsthm}
\usepackage{arydshln}
\usepackage[capitalize,noabbrev]{cleveref}

\usepackage{mathtools} 

\usepackage{amsmath,amsfonts,bm}

\def\1{\bm{1}}

\DeclareMathAlphabet{\mathsfit}{\encodingdefault}{\sfdefault}{m}{sl}
\SetMathAlphabet{\mathsfit}{bold}{\encodingdefault}{\sfdefault}{bx}{n}

\makeatletter
\renewcommand{\fnum@figure}{\textbf{Fig. \thefigure} | }
\renewcommand{\fnum@table}{\textbf{Table \thetable \ |}}

\makeatother
\usepackage[labelsep=space]{caption}
\usepackage[
singlelinecheck=false 
]{caption}

\usepackage{authblk}

\title{
Generative Design of Functional Metal Complexes Utilizing the Internal Knowledge of Large Language Models
}

\author[1, $ \dag $]{Jieyu Lu}
\author[1, $ \dag $]{Zhangde Song}
\author[1]{Qiyuan Zhao}
\author[2]{Yuanqi Du}
\author[1]{Yirui Cao}
\author[1, *]{Haojun Jia}
\author[1, $ \dag $, *]{Chenru Duan}
\affil[1]{Deep Principle Inc., Cambridge, MA, 02139}
\affil[2]{Department of Computer Science, Cornell University, Ithaca, NY, 14850}
\affil[$ \dag $]{These authors contribute equally}
\affil[*]{Corresponding to: haojunjia@deepprinciple.com, duanchenru@gmail.com}

\begin{document}

\maketitle

\begin{abstract}

The design of functional transition metal complexes (TMCs) is hindered by the combinatorial explosion of the search space spanned by various metals and ligands, necessitating efficient multi-objective optimization strategies. 
Traditional genetic algorithms (GAs) are frequently employed in this domain, utilizing random mutations and crossovers steered by explicit mathematical objective formulations to navigate the search space. 
The transfer and sharing of knowledge across different GA optimization tasks, however, remain challenging. 
Here, we introduce the integration of large language models (LLMs) into the evolutionary optimization framework (LLM-EO) for TMCs. 
LLM-EO significantly outperforms traditional GAs due to the intrinsic chemical knowledge embedded within LLMs, acquired during their extensive pretraining. 
Notably, without the need for supervised fine-tuning, LLMs can leverage the entirety of historical data amassed during the optimization processes, demonstrating superior performance compared to LLMs that are limited to the best TMCs identified in the evolutionary cycle. 
Specifically, LLM-EO identifies eight out of the top 20 TMCs with the largest HOMO-LUMO gaps by interrogating merely 200 candidates within a vast search space of 1.37 million TMCs.
Through prompt engineering using natural language, LLM-EO introduces unparalleled flexibility in multi-objective optimizations, especially when guided by seasoned researchers, thereby circumventing the necessity for intricate mathematical formulations. 
As generative models, LLMs possess the capability to propose novel ligands and TMCs with unique chemical properties by amalgamating both internal knowledge and external chemistry data, thus combining the benefits of efficient optimization and molecular generation.
With the increasing potential of LLMs, both in their capacity as pretrained foundational models and new strategies in post-training inference, we anticipate broad applications of LLM-based evolutionary optimization in the fields of chemistry and materials design.
\end{abstract}

\setstretch{1.8}


\section*{Introduction}
Evolutionary optimization (EO) has emerged as a powerful methodology for designing functional materials, facilitating efficient exploration of vast chemical spaces to identify candidates with desired properties \cite{Le2016May,Nandy2021Aug,Anstine2023Apr,WalshDD2024,DuNMI2024}. Drawing inspiration from biological evolution, EO algorithms incorporate operations such as selection, mutation, recombination, and reproduction to iteratively evolve solutions. 
Genetic algorithms (GAs) are a prominent subset of EO methods \cite{Goldberg1989Oct}, employing a fitness function to evaluate and rank the quality of each solution. 
These algorithms apply evolutionary operators to the highest-ranked candidates, iteratively enhancing solution quality through mutation and crossover until predefined performance criteria are achieved. 
Despite their simplicity, GAs have demonstrated effectiveness comparable to machine learning (ML) methods for certain molecular optimization tasks \cite{Jensen2019Mar,WenhaoMolBenchmark}. 
Moreover, the integration of GAs with ML techniques can markedly accelerate the fitness evaluation process, improving the overall efficiency of the exploration \cite{Janet2018Mar,Jennings2019Apr,Nigam2022,DuanJACSAu2023}. 
The adaptability and robustness of EO render it particularly suited for complex optimization challenges in chemistry, which are often characterized by large, multidimensional search spaces \cite{JensenTMC2024,GuzikScience2024}.

\qquad Designing chemical systems composed of multiple building blocks, such as transition metal complexes (TMCs), presents significant challenges\cite{Nandy2021Aug,MITScience2024}. 
The combinatorial explosion resulting from the diverse combinations of metals and ligands makes systematic exploration of the TMC chemical space difficult, surpassing the capabilities of current theoretical and experimental screening methods\cite{Nandy2021Aug,Wallach2024}. 
While exhaustive enumeration is impractical, efficiently navigating this vast chemical space is essential for discovering TMCs with optimized properties.
Furthermore, the need of multi-objective optimization, which is often necessary in the design of functional materials and catalysts, adds an additional layer of complexity to the design of fitness functions in standard GA\cite{Laplaza2022Jun}. 
Existing EO applications in TMC design have shown promise in optimizing properties such as catalytic activity and stability\cite{Nandy2021Aug,Strandgaard2024}. 
For instance, Kneiding et al. developed the Pareto-Lighthouse Multiobjective Genetic Algorithm (PL-MOGA), providing fine-grained control over multiple optimization objectives for optimizing properties of TMCs in a 1.37M chemical space\cite{Kneiding2024Apr}.
However, two factors limit the generative power of current EO approaches for TMCs: the inherent complexity of formulating multi-objective fitness functions and the trade-offs between operating on smaller ligand fragments and whole ligands. 
Operating on whole ligands is computationally efficient because it limits the chemical space, making exploration faster. 
In contrast, smaller fragments increase diversity by allowing more combinations and novel structures, but they expand the chemical space, making it computationally demanding to explore\cite{Seumer2024May}.
These challenges necessitate further methodological innovations to unlock the full potential of EO in designing new TMCs.

\qquad Recent advancements in large language models (LLMs), initially developed for natural language processing, have generated substantial interest in their application across a broad range of scientific disciplines. 
These models have shown remarkable potential in areas such as text mining \cite{Zheng2023Aug,Gupta2022May,Lee2020Feb,Peng2020May}, scientific planning \cite{Bran2023Apr,Boiko2023Dec}, reaction prediction \cite{Sagawa2023Nov,Lu2022Mar,Irwin2022Jan,Schwaller2019Sep}, molecular generation \cite{Kyro2024Feb,Fang2023Jan,Bagal2022May}, property prediction \cite{Liu2024Jun,Shen2024Feb,Ock2023Dec,Ross2022Dec,SmitLLMNature2024}, and educational applications \cite{Du2024Jun,Zhang2024Jun,Khan2023Mar,Tsai2023Jul,Hocky2022}. 
By integrating with evolutionary algorithms, LLMs have demonstrated enhanced capabilities, advancing the exploration and optimization processes in evolving concept libraries \cite{grayeli2024symbolic} and heuristic design \cite{Yao2024Sep}. 
Recently, Wang et al. employed LLMs for the design of small organic molecules via SMILES string generation, highlighting the potential of LLMs as a powerful tool for chemical discovery \cite{wang2024efficient}. 
Nevertheless, their applications in complex chemical systems or materials composed of multiple building blocks under multi-objective optimization have yet to be realized.

\qquad In this study, we introduce LLM-EO, an integration of LLMs into the framework of EO, and apply it to both single- and multi-objective optimization tasks for transition metal complexes (TMCs).
With only a few examples, LLMs can propose TMCs with desired properties by leveraging their extensive chemistry knowledge acquired during pretraining on a large corpus.
In addition, LLM-EO significantly outperforms GA in optimizing TMCs for a single objective. 
Its performance is further enhanced when all historical data from the optimization process is retained, offering application scenarios akin to closed-loop optimizations in a laboratory setting. 
By utilizing natural language instructions, LLM-EO demonstrates exceptional flexibility in multi-objective optimizations, reducing the need for complex mathematical formulations. 
Lastly, LLM-EO leverages the generative capabilities of LLMs to generate entirely new ligands and TMCs, significantly accelerating the optimization process. By exploring an boundless design space that goes beyond the conventional chemical spaces defined by chemical intuition, LLM-EO opens up limitless possibilities for innovative compound design.
Our results highlight LLM-EO as a continuously improvable framework for the design of functional materials.

\section*{Results}

\begin{figure*}[t!]
    \centering
    \includegraphics[width=0.98\textwidth]{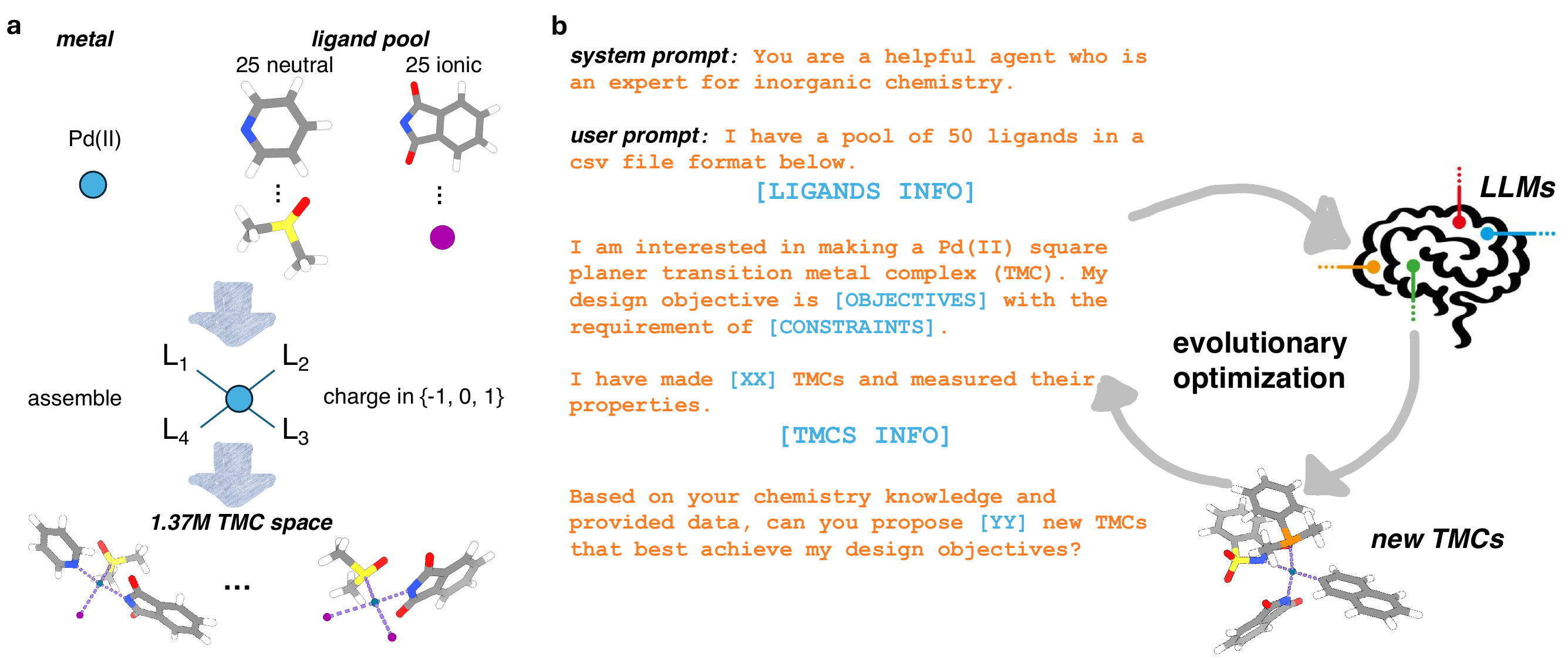}
    \caption{\textbf{Overview of the design space of TMCs and LLM-EO.}
    \textbf{a.} Construction of 1.37M space of square planar TMCs. The building blocks of TMCs consist of Pd in oxidation of II (i.e., metal) and a ligand pool with 25 neutral ligands and 25 ionic ligands. By assembling them in a square planar geometry with the constraint of total TMC charge being -1, 0, or 1, a space of 1.37M TMC is constructed.
    \textbf{b.} Workflow of LLM-EO. A prompt is
    engineered towards designing TMCs with both generic instructions (text in orange) and certain information, constraints, and objectives (text in blue). This prompt interacts with LLM (e.g., through API calls), which returns a new set of TMCs. These new TMCs are fed into the prompt with updated information, closing the loop of evolutionary optimization.
    Atoms are colored as follows: Pd for sky blue, C for gray, N for blue, O for red, P for orange, S for yellow, I for purple, and H for white.
    Purple dashed lines in TMCs represent dative bonds between Pd(II) and ligands.
    }
    \label{fig:overview}
    
\end{figure*}

\paragraph{Overview of LLM-EO.}
LLM-EO describes the integration of LLMs with EO, wherein potential candidates for exploration are proposed by the LLMs (see \textit{\nameref{method:LLMEO}}). 
Drawing inspiration from the principles of GA, we developed an iterative optimization workflow (Fig. \ref{fig:overview}b). 
In each iteration, a prompt is meticulously crafted, comprising generic instructions alongside specific information, constraints, and objectives, and is then presented to an LLM (see \textit{\nameref{method:PE}}). 
The LLM subsequently generates a new set of task-specific TMCs, which are evaluated and integrated into the prompt for the subsequent iteration, effectively completing the EO cycle. 
To evaluate the intrinsic chemical knowledge acquired by LLMs during pretraining, we utilize commercially available LLM checkpoints without applying explicit supervised fine tuning on any TMCs (Supplementary Table TA).

\qquad Despite its widespread application across various physical science disciplines, we examine the potential of LLM-EO in single- and multi-objective optimization tasks. 
Specifically, we utilize a chemical space comprising 1.37 million TMCs from Kneiding et al.\cite{Kneiding2024Apr} to evaluate the performance of LLM-EO (see Fig. \ref{fig:overview}a and \textit{\nameref{method:1.37MSpace}}). 
This dataset includes Pd(II) square planar complexes constructed using four monodentate ligands chosen from a pool of 50 ligands, which consists of 25 neutral and 25 monoanionic ligands, with the added constraint that the total charge of a TMC must be -1, 0, or 1.
Two properties of these 1.37M TMCs, i.e., HOMO-LUMO gap and polarisability, are computed using molSimplify\cite{molSimplify} and GFN2-xTB\cite{grimme2017robust,bannwarth2019gfn2} (see \textit{\nameref{method:EvalTMC}}). 
Compared to conventional EO algorithms such as GA, LLM-EO provides three key advantages:
First, it offers flexible objective control, enabling the articulation of more adaptable and nuanced optimization objectives in natural language as opposed to complex mathematical formulations.
Second, it enhances optimization efficiency through the rich chemical knowledge embedded in LLMs and their ability to learn from few-shot samples.
Third, it facilitates the generative design of new chemistry.
The inherent generative capabilities of LLMs allow LLM-EO not only to efficiently navigate a predefined search space but also to creatively generate novel TMCs.
These characteristics position LLM-EO as an emerging and powerful tool for chemical design.

\begin{figure*}[t!]
    \centering
    \includegraphics[width=0.86\textwidth]{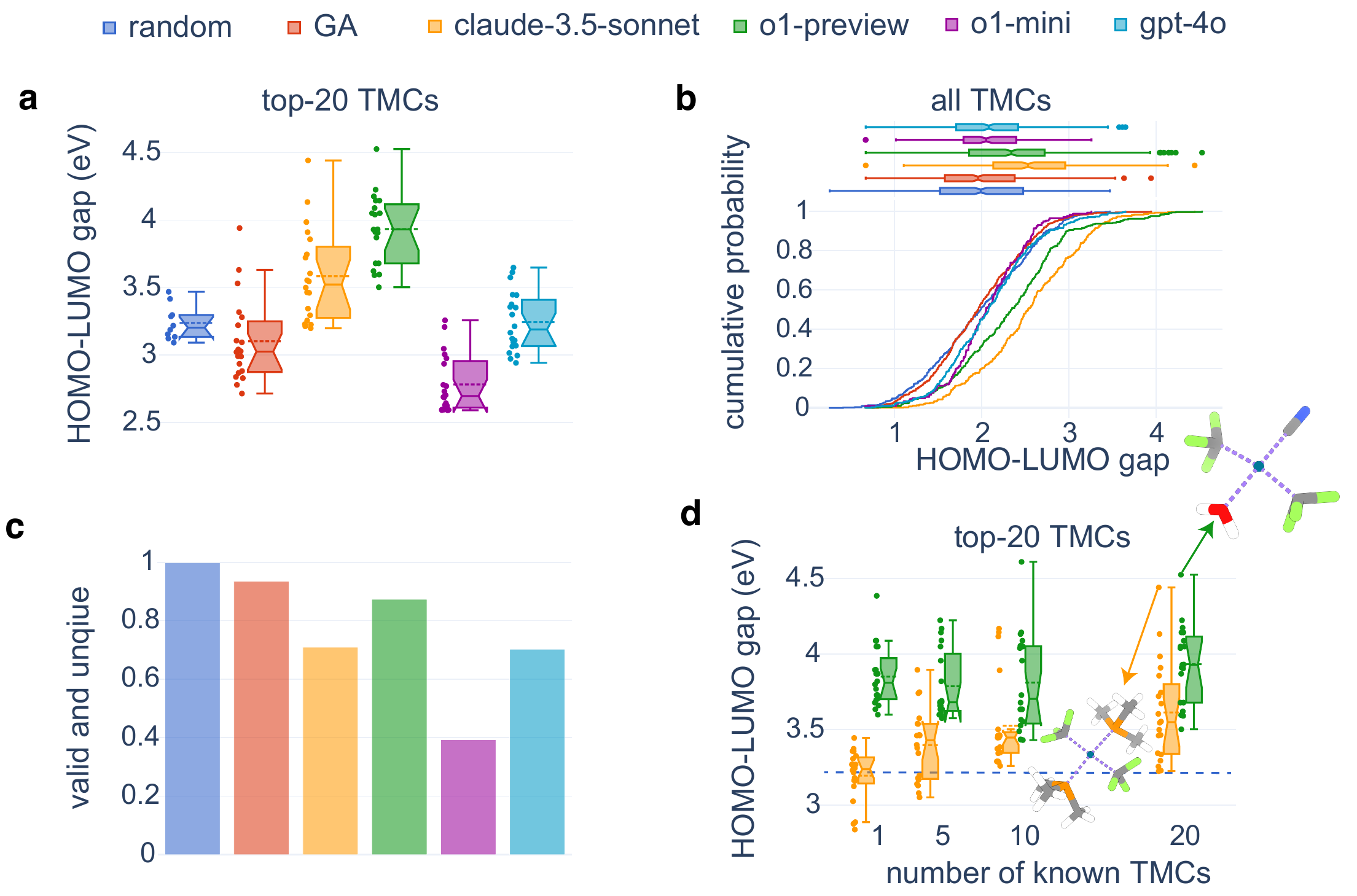}
    \caption{\textbf{Proposing new TMCs with few-shot LLMs.}
    \textbf{a.} Box plot (notched) for the distribution of HOMO-LUMO gap of top-20 TMCs among the 200 TMCs proposed by different approaches, providing 20 TMCs as known. Random is for blue, GA for red, claude-3.5-sonnet for orange, o1-preview for green, o1-mini for purple, and gpt-4o for sky blue.
    \textbf{b.} Cumulative probability of HOMO-LUMO gaps of all TMCs proposed by various approaches and their overall distribution (box plot at top margin). 
    \textbf{c.} Fraction of both valid and unique TMCs in 200 TMC proposals.
    \textbf{d.} Box plot (notched) for the distribution of HOMO-LUMO gap of top-20 TMCs among the 200 TMCs proposed, measured with a varying number of known TMCs provided in the prompt. Only the results for claude-3.5-sonnet (orange), o1-preview (green) are shown. The TMC with largest HOMO-LUMO gap for claude-3.5-sonnet and o1-preview is shown, correspondingly. The mean of random top-20 TMCs is shown with a blue dashed line.
    Atoms are colored as follows: Pd for sky blue, C for gray, N for blue, O for red, P for purple, F for green, and H for white.
    Purple dashed lines in TMCs represent dative bonds between Pd(II) and ligands.
    }
    \label{fig:few_shots}
\end{figure*}

\paragraph{LLMs proposing high-quality TMCs in few-shot learning.}
An important question to address is whether current LLMs have acquired chemical knowledge during their pretraining on extensive corpora, and if so, whether they can utilize this knowledge to derive meaningful chemical designs. 
Here, we focus on a challenging task of designing TMCs with maximized HOMO-LUMO gaps for single-objective optimization, which represent the energy difference between the highest occupied molecular orbital (HOMO) and the lowest unoccupied molecular orbital (LUMO), a complex task due to the intricate electronic structures involved.
To evaluate the ability of LLMs to learn structure-property relationships from limited data and to propose novel TMC designs within a constrained chemical space, we provide the models with 50 ligands represented in SMILES notation, along with 20 randomly sampled initial TMCs from the 1.37M TMC space, each accompanied by their corresponding HOMO-LUMO gaps (see Fig. \ref{fig:overview}b, \textit{\nameref{method:1.37MSpace}}, and Supplementary Text SA).
To demonstrate the current upper limit of LLM-EO, we employ the best commercially available LLMs from Anthropic (claude-3.5-sonnet) and OpenAI (o1-preview) within the LLM-EO framework (see \textit{\nameref{method:PE}}). 
According to OpenAI, o1-preview represents a marriage between strong pretrained foundational models with post-training inference to enhance reasoning capabilities \cite{o1Report}. 
For comparative analysis, we also include gpt-4o, a top-performing foundational model that does not prioritize reasoning, and o1-mini, a smaller foundational model utilizing the same post-training inference techniques as o1-preview does.
Throughout this work, results from random selection and GA are used as baseline (see \textit{\nameref{method:GA}}).

\qquad With few-shot learning, the top-20 TMCs proposed by claude-3.5-sonnet and o1-preview exhibit significantly larger HOMO-LUMO gaps compared to those obtained by random selection and GA (Fig. \ref{fig:few_shots}a). 
With only 400 proposals, both claude-3.5-sonnet and o1-preview identify candidate TMCs with HOMO-LUMO gaps greater than 4.45 eV, which places them in the top 0.002 \% (32 TMCs in total) of the 1.37M TMC space, underscoring their ability to pinpoint highly promising candidates for design objectives. 
For instance, both models recognize that ligands with strong electron-withdrawing effects (e.g., CF$_3^-$) and strong-field ligands (e.g., CN$^-$) contribute to large HOMO-LUMO gaps in square planar Pd(II) TMCs (Fig. \ref{fig:few_shots}d). 
Additionally, the overall distribution of TMCs proposed by claude-3.5-sonnet and o1-preview shifts towards larger HOMO-LUMO gaps compared to baseline methods (Fig. \ref{fig:few_shots}b).
Despite the numerous constraints on the design space and the limited examples provided during few-shot learning, both claude-3.5-sonnet and o1-preview achieve a high rate of validity and uniqueness, exceeding 71 \% (Fig. \ref{fig:few_shots}c). 
In contrast, o1-mini and gpt-4o perform significantly worse across all evaluation metrics compared to o1-preview (Fig. \ref{fig:few_shots}a-c). 
This disparity is attributed to either a lack of chemical knowledge due to the smaller size of the foundation model or insufficient reasoning capabilities during post-training inference. 
These observations highlight the model-dependent nature of LLM-EO, while also indicating its potential for sustainable improvement, which would benefit from the scaling law for foundation model pretraining and recent advancements in post-training inferences, such as chain-of-thought reasoning \cite{CoT} and self-play reinforcement learning\cite{AlphaGoZero,SelfPlayRL}. 

\qquad To investigate the influence of the number of initial TMCs provided in a few-shot learning setup, we vary the quantity of known TMCs included in the prompt to assess the resulting quality of the top-performing TMCs generated by LLM-EO. 
As one may expect, we observe that increasing the number of known initial TMCs enhances the HOMO-LUMO gaps of top-20 TMCs found by both claude-3.5-sonnet and o1-preview models (Fig. \ref{fig:few_shots}d). 
However, the improvement is more pronounced in claude-3.5-sonnet, possibly due to the exceptional performance of o1-preview with very limited known TMCs.
Notably, despite only a single TMC is provided, o1-preview achieves superior performance, demonstrating a 1 eV larger HOMO-LUMO gap in the top-20 TMCs compared to the baseline of random selection.
These findings indicate that LLMs can yield promising results even with limited data due to the chemical knowledge acquired during pretraining, offering a valuable balance between data requirements and performance.

\begin{figure*}[t!]
    \centering
    \includegraphics[width=0.98\textwidth]{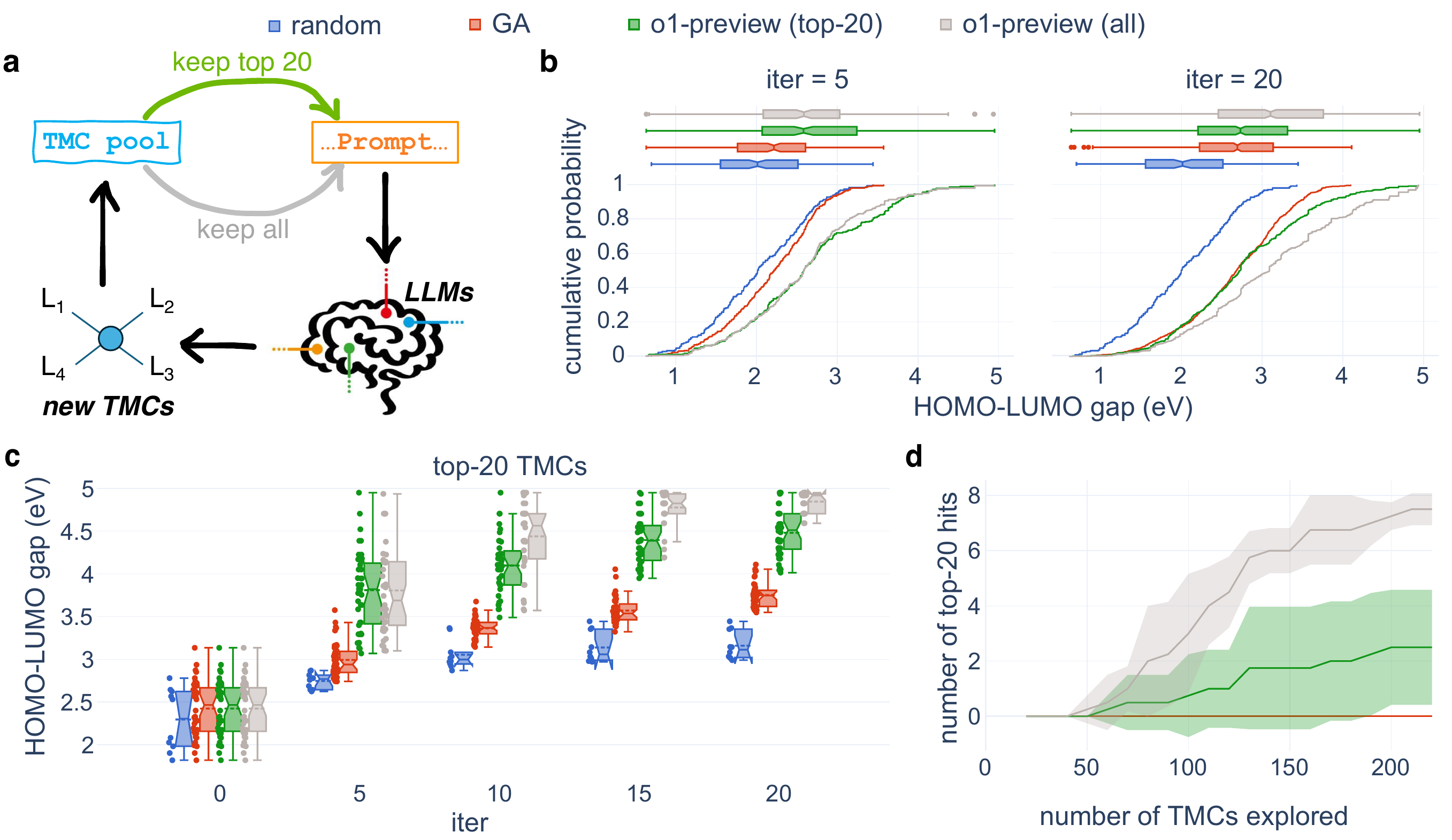}
    \caption{\textbf{Maximizing HOMO-LUMO gap with LLM-EO.}
    \textbf{a.} Schematic of LLM-EO, with the option of keeping top-20 TMCs with largest HOMO-LUMO gaps in the prompt (green) or keeping all historical data (gray) during the evolutionary optimization.
    \textbf{b.} Cumulative probability of HOMO-LUMO gaps of all TMCs proposed by various approaches at iteration = 5 (left) and iteration = 20 (right) during the evolutionary optimization. Random sampling is used as blue, GA as red, o1-preview with only top-20 TMCs kept in green, and o1-preview with all historical data in gray. The overall distribution of each approach is shown as box plot at top margin. 
    \textbf{c.} Box plot (notched) for the distribution of HOMO-LUMO gap of top-20 TMCs among the 200 TMCs proposed during the evolutionary optimization, providing 20 random TMCs as initial samples.
    \textbf{d.} Number of hits as top-20 TMCs with largest HOMO-LUMO gap among the 1.37M space versus the number of TMCs proposed during the evolutionary optimization. Solid lines are the average of three independent runs at different random seed, and their shedding corresponds to the standard deviation.
    }
    \label{fig:gap_eo}
\end{figure*}

\paragraph{Efficient optimization of HOMO-LUMO gap with LLM-EO.}
Upon confirming that LLMs can propose TMCs capable of achieving specific objectives using their inherent chemical knowledge, we progress towards closing the EO loop by integrating the 10 newly proposed TMCs with the original set (Fig. \ref{fig:gap_eo}a and Supplementary Text SB). 
Initially, a retention strategy operates concurrently with GA, where only TMCs exhibiting the highest fitness values—specifically, the top-20 largest HOMO-LUMO gaps—are retained in the prompt for subsequent iterations (see \textit{\nameref{method:LLMEO}}).
Given that the o1-preview model yields the best results in few-shot learning, we choose to utilize o1-preview as the core model for LLM-EO throughout the remainder of this work. 
At an early stage of optimization, with only 50 TMC proposals, LLM-EO demonstrates its efficacy in identifying TMCs with significant HOMO-LUMO gaps, outpacing GA, which has not yet shown substantial improvement over random selection (Fig. \ref{fig:gap_eo}b).
As the number of iterations increases, GA also begins to show advantages over random selection; however, it continues to lag behind LLM-EO. 
Similar trends are observed when focusing on the top-20 TMCs found truncated at each iteration, where LLM-EO consistently outperforms GA and random selection (Fig. \ref{fig:gap_eo}c).
The improvement on HOMO-LUMO gaps for top-performing TMCs is much more significant compared to the overall distribution, suggesting the exceptional efficiency of LLM-EO on identifying TMCs extreme properties, a feature that is often desired in materials optimization. 

\qquad In contrast to GA where only top-performing candidates are retained, there is a natural desire to utilize all historical data during the optimization process. 
Consequently, we adopt a different retention strategy than GA, wherein all previously evaluated TMCs are preserved, irrespective of their HOMO-LUMO gaps (see Fig. \ref{fig:gap_eo}a and \textit{\nameref{method:LLMEO}}).
At early iterations, the distinction between retaining only top-performing and all TMCs is minimal, owing to the limited number of TMCs explored (Fig. \ref{fig:gap_eo}b and c). 
However, as more TMCs are assessed during optimization, we observe significant benefits in maintaining records of all TMCs, which not only enhances the overall distribution of HOMO-LUMO gaps but also bolsters the identification of the top-20 TMCs.
These findings underscore the capacity of LLMs to learn from an increasing volume of evaluations, even when certain experiments may be deemed unsuccessful due to low fitness levels.

\qquad By leveraging historical data during optimization, LLM-EO efficiently identifies top-tier candidates within a vast design space. 
Notably, by retaining all historical data, LLM-EO consistently identifies 8 out of the top-20 TMCs with the largest HOMO-LUMO gaps, despite evaluating only 200 TMCs—a mere 0.015\% of the 1.37M design space (Fig. \ref{fig:gap_eo}d). 
In addition, 11.5\% (equating to 23) of the 200 TMCs proposed by LLM-EO rank among the top-200 TMCs with the largest HOMO-LUMO gaps within the 1.37M space (Supplementary Figure FA). 
In stark contrast, neither GA nor random selection yields any TMCs that fall into the top-200 of the 1.37M space. 
The ability of LLMs to harness accumulated knowledge positions LLM-EO as an exceptional laboratory assistant, which minimizes the time and effort needed to discover functional metal complexes by assimilating comprehensive historical data.

\begin{figure*}[t!]
    \centering
    \includegraphics[width=0.98\textwidth]{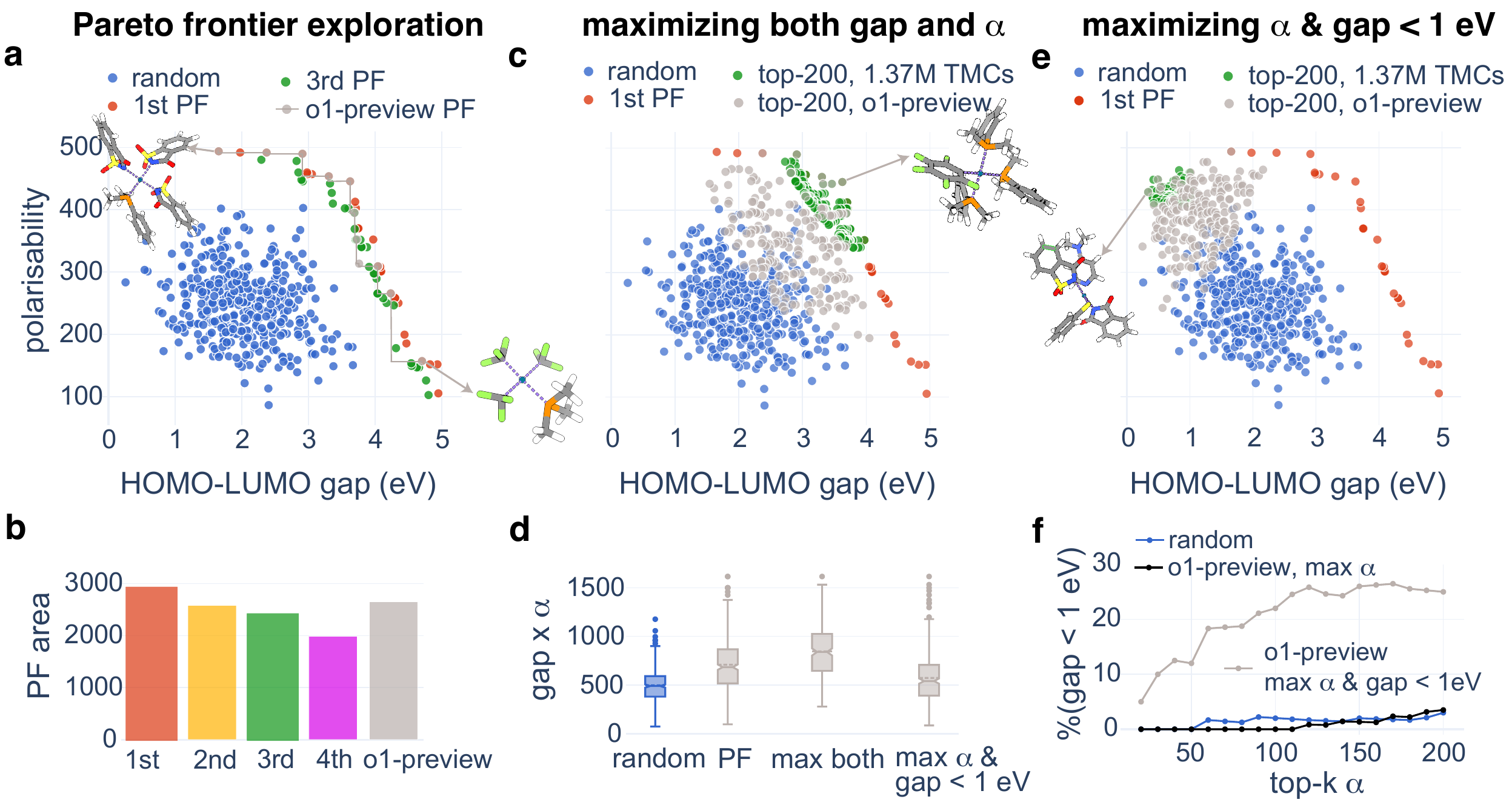}
    \caption{\textbf{Multi-objective optimization with LLM-EO} for Pareto frontier exploration (left), maximizing both HOMO-LUMO gap and polarisability ($\alpha$, middle), and maximizing HOMO-LUMO gap while keeping $\alpha$ < 1 eV.
    \textbf{a.} Polarisability versus HOMO-LUMO gap for random 400 TMCs (blue). The Pareto frontier (PF) for the 1.37 TMC space (red), the 3rd PF for the 1.37M TMC space (green), and the PF found by o1-preview with 400 LLM-EO exploration (gray), are shown. A solid step-wise line is drawn to shown the frontier area obtained by o1-preview in 400 TMCs exploration.
    \textbf{b.} Area under PF curves for the true 1st PF (red), 2rd PF (orange), 3rd PF (green), 4th PF (purple) for the 1.37M TMC space, and PF found by o1-preview (gray).
    \textbf{c.} Polarisability versus HOMO-LUMO gap for random 400 TMCs (blue) and top-200 TMCs with largest multiplication of HOMO-LUMO gap and polarisability for the 1.37M space (green) and 400 TMCs explored by o1-preview (gray). The Pareto frontier (PF) for the 1.37M TMC space (red) is also shown.
    \textbf{d.} Box plot (notched) for the distribution of HOMO-LUMO gap and polarisability multiplication of random TMCs (blue) and TMCs proposed by o1-preview (gray) during PF exploration, maximizing both gap and polarisability, and maximizing polarisability while keeping gap < 1 eV.
    \textbf{e.} Polarisability versus HOMO-LUMO gap for random 400 TMCs (blue) and top-200 TMCs with largest polarisability while gap < 1 eV for the 1.37M space (green) and 400 TMCs explored by o1-preview (gray). The Pareto frontier (PF) for the 1.37 TMC space (red) is also shown.
    \textbf{f.} Percentage of TMCs with HOMO-LUMO gap < 1 eV at various top-k polarisability by random sampling (blue), o1-preview at single-objective optimization for only maximizing polarisability (black), and o1-preview at multi-objective optimization for maximizing polarisability while requiring HOMO-LUMO gap < 1 eV (gray).
    Select TMCs are shown for each case as insets.
    Atoms are colored as follows: Pd for sky blue, C for gray, N for blue, O for red, P for orange, F for green, and H for white.
    Purple dashed lines in TMCs represent dative bonds between Pd(II) and ligands.
    }
    \label{fig:2d_eo}
\end{figure*}

\paragraph{Flexible multi-objective optimization guided by natural language.}
Unlike traditional EO algorithms, such as GA, which require explicit mathematical formulations for multi-objective optimization, the LLM-EO approach utilizes the flexibility of natural language instructions. 
We illustrate this adaptability through three distinct multi-objective (both targeting on HOMO-LUMO gap and polarisability) optimization tasks within the 1.37M TMC space, where all historical data are retained during the execution of LLM-EO with o1-preview: First, Pareto frontier (PF) optimization, second, maximizing both the HOMO-LUMO gap and polarisability, and lastly, maximizing polarisability with a constraint that the HOMO-LUMO gap remains below 1 eV (see \textit{\nameref{method:LLMEO}}).

\qquad The PF represents the set of optimal solutions in which enhancement in one objective requires a trade-off in another. 
With only 400 proposals, LLM-EO identifies nine TMCs that reside on the true (that is, 1st) PF of the 1.37M space (Fig. \ref{fig:2d_eo}a and Supplementary Text SC). 
The PF established by LLM-EO successfully identifies TMCs with electron-withdrawing ligands, such as Pd(II)(CF$_3^-)_3$(P(CH$_3)_3$), which maximize the HOMO-LUMO gap, as well as those with highly polarized ligands, such as 3-Oxo-1,2-benzisothiazol-2-ide 1,1-dioxide (C$_7$H$_4$NO$_3$S$^-$), which maximizes polarisability.
All TMCs located on the PF found by LLM-EO within 400 proposals surpass the 3rd PF of the 1.37M space (Fig. \ref{fig:2d_eo}a).
The area under the curve (AUC) can be utilized to assess the positioning of a PF in the solution space, where a larger AUC indicates more effective trade-offs among multiple objectives. 
By exploring only 0.03 \% of the total space, the AUC of LLM-EO's PF ranks already between that of the 1st and 2nd PFs of the 1.37M space (Fig. \ref{fig:2d_eo}b).
This highlights the efficacy of LLM-EO in identifying top-performing candidate TMCs within a multi-dimensional space.

\qquad When tasked with maximizing both the HOMO-LUMO gap and polarisability, LLM-EO effectively proposes TMCs concentrated in the top-right quadrant of the 1.37M space, illustrating its capability to target both objectives simultaneously (Fig. \ref{fig:2d_eo}c and Supplementary Text SD). 
By evaluating only 400 TMCs, LLM-EO identifies 18 TMCs that overlap with the top-200 theoretically optimal TMCs, judged by the product of the HOMO-LUMO gap and polarisability. 
Among these, Pd(II)(P(CH$_3$)$_2$(C$_6$H$_5$))$_3$(C$_6$F$_5$)$^-$ emerges as the TMC with the highest score (product of two properties), achieving a balanced emphasis on the HOMO-LUMO gap and polarisability by incorporating a polar C$_6$H$_5$ group on the phosphorous ligand (Fig. \ref{fig:2d_eo}c). 
In contrast to random selection and PF exploration, LLM-EO distinctly produces TMCs with an overall higher product of the HOMO-LUMO gap and polarisability when tasked with maximizing both properties, demonstrating the effectiveness of prompt instruction (Fig. \ref{fig:2d_eo}d). 

\qquad Finally, we assess the effectiveness of incorporating constraints within the LLM-EO for multi-objective optimization, a scenario frequently encountered in practical functional materials design. 
Specifically, we task the LLM-EO with maximizing the polarisability of a TMC while maintaining a very small HOMO-LUMO gap of less than 1 eV, where only 3.8 \% of TMCs in the 1.37M space meets this criterion (Supplementary Figure FB and Supplementary Text SE).
With these new instructions in the prompt, the LLM-EO predominantly explores TMCs in the top-left corner of the 1.37M space, identifying 19 of the top-200 TMCs with the largest polarisability that also satisfy the HOMO-LUMO gap constraints within 400 proposals (Fig. \ref{fig:2d_eo}e). 
Among these LLM-EO finds a heteroleptic TMC with four distinct ligands, achieving a extremely small HOMO-LUMO gap of 0.3 eV and a high polarisability of 430 a.u.
The products of HOMO-LUMO gap and polarisability for the 400 TMCs proposed by LLM-EO are naturally lower than those from random selection due to the constraints imposed on HOMO-LUMO gap (Fig. \ref{fig:2d_eo}d). 
Since TMCs with the highest polarisability do not inherently have small HOMO-LUMO gaps, it is exceedingly rare (< 2 \%) for them to achieve a HOMO-LUMO gap under 1 eV, as demonstrated by both random selection and single-objective optimization for polarisability using LLM-EO (Fig. \ref{fig:2d_eo}f).
Incorporating the HOMO-LUMO gap constraint into the prompt, however, significantly increases the number of TMCs with a HOMO-LUMO gap smaller than 1 eV while maintaining top-tier polarisability. 
This result highlights the potential of LLM-EO to accurately control and optimize desired properties through natural language instructions. 
Such precise regulation eliminates the necessity for complex scoring function formulations in multi-objective optimizations, thus enabling efficient exploration of diverse chemical spaces with targeted property combinations.

\begin{figure*}[t!]
    \centering
    \includegraphics[width=0.98\textwidth]{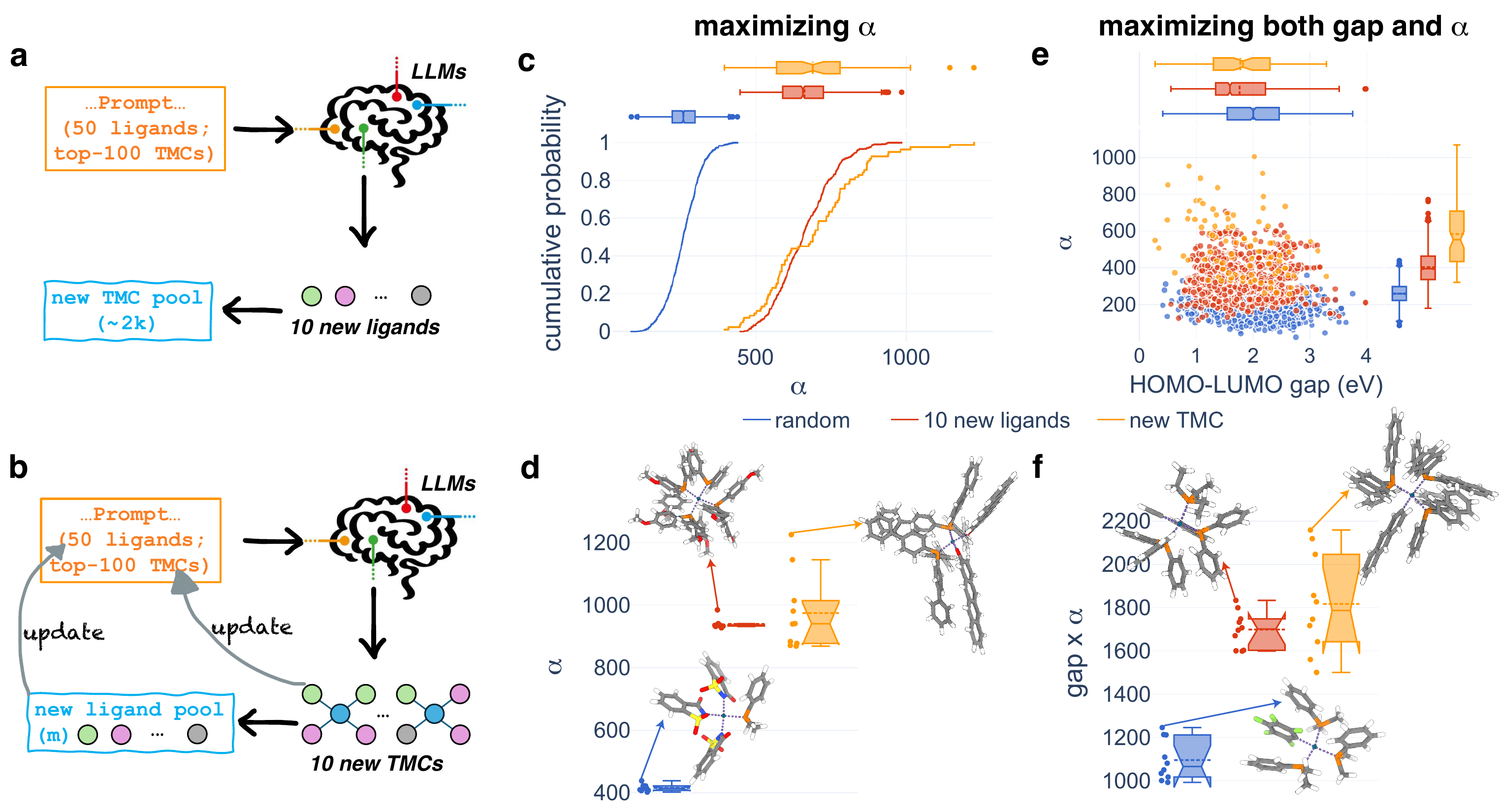}
    \caption{\textbf{Generating new ligands and TMCs with LLMs.} 
    \textbf{a.} Workflow for assembling new TMC space by asking LLMs to propose new ligands outside of the original ligand pool.
    \textbf{b.} Workflow for direct iterative TMC optimization by asking LLMs to propose new TMCs that are made by new ligands outside of the original ligand pool.
    \textbf{c.} Cumulative probability of polarisability of TMCs proposed by various approaches. Random sampling is used as blue, using 10 ligands generated by o1-preview in one shot in red, LLM-EO by o1-preview for generating new TMCs iteratively in orange.
    \textbf{d.} Box plot (notched) for the distribution of polarisability for TMCs explored by various approaches. The best TMC for each approach is shown.
    \textbf{e.} Polarisability versus HOMO-LUMO gap for random 400 TMCs (blue), TMCs constructed by 10 ligands generated by o1-preview in one shot (red), and TMCs generated by LLM-EO using o1-preview iteratively (orange). Box plots for the distribution of each property are shown at margins.
    \textbf{f.} Box plot (notched) for the distribution of the product of HOMO-LUMO-gap and polarisability (that is, $\alpha$) for TMCs explored by various approaches. The best TMC for each approach is shown.
    Atoms are colored as follows: Pd for sky blue, C for gray, N for blue, O for red, F for green, P for orange, S for yellow, and H for white.
    Purple dashed lines in TMCs represent dative bonds between Pd(II) and ligands.
    }
    \label{fig:gen_tmc}
\end{figure*}

\paragraph{Utilizing the generative power of LLMs for \textit{de novo} TMC design.}
To fully harness the generative capabilities of LLMs, we delve deeper into the potential of LLM-EO for designing novel ligands and TMCs beyond a predefined chemical space limited by an initial ligand pool (Supplementary Figure FC). 
We employ two distinct strategies for this purpose:
First, one-shot generation approach for novel ligands (Supplementary Text SF). 
By leveraging the top-performing TMCs and the initial set of 50 ligands, the o1-preview model is utilized to generate a new pool of 10 ligands in a single shot (see Fig. \ref{fig:gen_tmc}a). 
These newly generated ligands are subsequently used to construct a chemical space comprising 2,200 square planar TMCs through enumeration (see \textit{\nameref{method:LLMEO}}). 
Lastly, HOMO-LUMO gaps and polarizabilities of these TMCs are calculated using molSimplify and GFN2-xTB (see \textit{\nameref{method:EvalTMC}}).
Second, iterative generation for novel TMCs (Supplementary Text SG). 
Similarly to the first approach, we start with the top-performing TMCs and the original set of 50 ligands. 
The o1-preview model generates 10 new TMCs, each incorporating at least one ligand not present in the initial ligand pool (see Fig. \ref{fig:gen_tmc}b). 
These 10 new TMCs are subsequently decomposed to extract a set of novel ligands generated by the o1-preview model as by-products. 
Finally, both the newly generated TMCs and ligands are integrated into the prompt for subsequent iterations of LLM-EO exploration (see \textit{\nameref{method:LLMEO}}).

\qquad We initially test two generative LLM-EO approaches on enhancing the polarisability of Pd(II)-based TMCs in a square planar geometry. 
Utilizing the 10 newly generated ligands from the o1-preview model, we synthesize 2,200 novel TMCs, which demonstrate substantially higher polarisability compared to those in the existing 1.37M TMC space, with a minimum polarisability recorded at 447 a.u. (Fig. \ref{fig:gen_tmc}c). 
These newly generated ligands are often bulky, featuring extended conjugated systems or electron-rich moieties—a trait consistent with the chemical intuition that larger and more diffuse electron clouds enhance polarisability (Supplementary Figure FD).
By iteratively generating new TMCs, the LLM-EO approach is also able to identify complexes with significantly greater polarisability (Fig. \ref{fig:gen_tmc}c). 
When examining the top-performing 10 TMCs, both methods produced TMCs with polarisability values twice as high as the best TMCs found in the 1.37M space (Fig. \ref{fig:gen_tmc}d). 
Moreover, direct iterative generation via LLM-EO identifies a TMC with an exceptionally high polarisability exceeding 1,200 a.u., characterized by all four ligands containing extended conjugated rings and engaging in five subtle $\pi-\pi$ interactions (Fig. \ref{fig:gen_tmc}d).
This enhanced polarisability illustrates the generative capacity of LLM-EO in exploring chemical spaces that are typically inaccessible through conventional single-objective optimization methods. 

\qquad To further assess the limits of LLM-EO, we subjecte it to a multi-objective optimization task aimed at simultaneously maximizing polarisability and the HOMO-LUMO gap. 
Both approaches of LLM-EO—generating new ligands in a single step and the iterative generation of TMCs—successfully identify TMCs that significantly exceed the boundaries of the 1.37M space (Fig. \ref{fig:2d_eo}e). 
However, both methods exhibit a tendency to prioritize optimizing polarisability over the HOMO-LUMO gap. 
This may be due to the inherent constraints on the HOMO-LUMO gap of a Pd(II) square planar TMC, which is typically limited to approximately 5 eV, making it more challenging to extend in comparison to polarisability \cite{tmQM}. 
Nonetheless, with the iterative approach, LLM-EO is capable of generating TMCs that achieve a balanced effect between ligand field strength and conjugation. 
This results in TMCs with relatively large HOMO-LUMO gaps (> 3 eV) while maintaining significant polarisability (> 600 a.u.) (Fig. \ref{fig:2d_eo}f).
By harnessing the inherent generative nature of LLM-EO, TMCs with exceptional properties can be identified efficiently with only a few hundred trials. 
This demonstrates a significant improvement in efficacy compared to the conventional forward design approach, which involves screening a large, predefined chemical space.

\section*{Discussion}
Efficient navigation through vast chemical spaces is crucial for designing functional materials, particularly those composed of multiple building blocks, such as TMCs. 
We developed LLM-EO, a workflow that leverages the predictive and generative capabilities of LLMs for evolutionary optimization.
We demonstrated the effectiveness of LLM-EO in both single- and multi-objective optimization tasks for TMCs. 
The inherent generative abilities of LLMs enable LLM-EO to undertake \textit{de novo} design, creating TMCs that excel in their design objectives, thus surpassing the limitations of predefined chemical spaces.

\qquad We observe significant variation in the performance of LLM-EO when utilizing different LLMs. 
Both the quality of the pretrained foundation model and the mechanism of post-training inference are crucial factors, as exemplified by a case where o1-preview substantially outperforms gpt-4o and o1-mini in optimizing the HOMO-LUMO gap of TMCs with LLM-EO. 
Interestingly, claude-3.5-sonnet demonstrates superior performance in a similar context compared to gpt-4o, suggesting its potential to surpass o1-preview once post-training inference mechanisms are enhanced for better reasoning capabilities. 
Moreover, the consistently improved performance of LLM-EO with each new release of LLMs indicates an almost limitless potential for performance enhancement in the LLM-EO workflow for chemical design. This is particularly promising when strategically incorporating scientific literature into the corpus during the pretraining of an LLM.

\qquad The inherent characteristics of LLMs are prominently exhibited in the LLM-EO framework. 
For instance, retaining all historical data during the optimization process is more effective than merely preserving the top-performing candidates, as seen in genetic algorithms (GA). 
This efficacy is attributed to the capability of LLMs to efficiently learn from expansive contexts. 
Furthermore, LLM-EO allows for instruction using natural language, eliminating the necessity for explicit mathematical task formulations. 
This feature provides unparalleled flexibility for addressing complex multi-objective optimization challenges, such as those encountered in practical materials design, where multiple metrics need to be satisfied. 
Finally, due to its generative nature, LLM-EO can discover entirely new materials possessing exceptional properties that are not present within a predefined chemical space, which might typically rely on intuition or enumeration. 

\qquad Within LLM-EO framework, the two models, o1-preview and claude-3.5-sonnet, demonstrate significantly better performance compared to open-source models and those developed in certain regions in our internal tests. 
Despite we do not aim this work as a benchmark study, this difference highlights potential challenges in ensuring equity in scientific progress toward artificial general intelligence, particularly as some organizations may focus on ambitious, revenue-oriented product development strategies.
Although we demonstrate the utility of LLM-EO in optimizing TMCs for HOMO-LUMO gaps and polarisability, it holds promise for numerous applications across biology, chemistry, and materials science, including DNA sequence design, the discovery of both homogeneous and heterogeneous catalysts, and the construction of functional materials with multiple building blocks, such as metal-organic frameworks. 
As LLM capabilities continue to expand, we anticipate that the LLM-EO framework introduced in this work will become increasingly beneficial for other domains in scientific discovery.

\section*{Methods}

\paragraph{Genetic algorithms.}\label{method:GA}
Genetic algorithms (GAs) are optimization techniques inspired by natural selection and genetics, effective in constructing systems from components like TMCs. 
In this work, GA is used as the baseline for the performance in proposing new TMCs.
The GA process begins with the same initial population sampled from the 1.37M compound space. HOMO-LUMO gap values were obtained from PL-MOGA\cite{Kneiding2024Apr}, calculated using the GFN2-xTB method. 
Then an evolutional optimization (EO) process iteratively updates the population through genetic operations, namely crossover and mutation, to optimize the objective function:
\begin{enumerate}
  \item \textit{Crossover.} This operation involved combining parts from two parent solutions to produce offspring.
  In our experiment, pairs of TMCs were selected and their ligands exchanged to create new candidate complexes.
  \item \textit{Mutation.} After crossover, mutation introduced random alterations to the offspring's chemical structure. Specifically, one or more ligands from parental TMCs were replaced with random ligands from a library of 50 ligands.
\end{enumerate}

\paragraph{Implementation Details of LLM-EO.}\label{method:LLMEO}
In this study, the intrinsic chemical knowledge of Large Language Models (LLMs) and their few-shot learning capabilities are leveraged within an LLM-based evolutionary optimization framework (LLM-EO). At each iteration, new TMCs are proposed by the LLM, subject to specific constraints. This process is guided by design objectives articulated in natural language, obviating the need for explicit mathematical formulations. The proposed TMCs are then integrated into the prompt, serving as the knowledge base for subsequent TMC proposals in the next iteration.
This LLM-EO approach differs from traditional evolutionary optimization algorithms (EO) such as Genetic Algorithms (GA), where operations are limited to a fixed set of manipulations like mutation and crossover. Furthermore, unlike Bayesian optimization\cite{DuanJACSAu2023}, evaluation of the fitness function across the entire design space at each iteration is not required. Several LLM-EO configurations were explored in this work, and are detailed below.

\textit{Direct few-shot proposal. }
The process begins with a pool of 50 ligands, each characterized by its structure, ligand ID, and connecting atom information as part of generic instructions. Additionally, 20 initial TMCs are provided within the prompt to guide a LLM (e.g., o1-preview). The LLM is asked to propose 10 new TMCs that optimize a given objective (e.g., maximizing the HOMO-LUMO gap). This request is repeated 20 times with the same set of known TMCs, resulting in a total of 200 LLM-proposed TMCs. To mitigate bias introduced by the initial TMC sampling, two different random seeds are used for this process.

\textit{Evolutionary optimizations with LLMs. }
Starting with 20 initial TMCs, an LLM (e.g., o1-preview) is then employed to propose 10 new TMCs. These new TMCs are merged with the existing data. From this combined pool, the top-20 candidates with the highest fitness values are selected as the known TMCs for the next iteration. This iterative process is repeated for 20 iterations in single-property optimization tasks (Fig. \ref{fig:gap_eo}) and 40 iterations in multi-objective optimization tasks (Fig. \ref{fig:2d_eo}). Similarly, to minimize bias from the initial TMC sampling, two different random seeds are used to sample the initial known TMCs.

\textit{Utilizing all historical records. }
Selecting the top-k samples in traditional evolutionary optimization requires explicit mathematical formulation of objectives, which can be challenging for complex multi-objective tasks. Therefore, we further explore a setup where all historical data are preserved during the evolutionary optimization using LLM. Specifically, all TMCs explored during the optimization process, regardless of their fitness values, are included in the prompt for the next iteration. This mimics the human learning experience where one can learn from both good and bad examples.
To prevent the LLM from overemphasizing specific records, the historical TMC data within the prompt is randomly shuffled before each iteration. This randomization enhances sampling diversity by presenting the LLM with different prompt orderings while maintaining the same underlying set of known TMCs.

\textit{Flexible 2D optimizations with pure natural language. }
LLM-EO can be readily extended to multi-objective optimizations by tuning the languages describing the task, circumventing the necessity for explicit mathematical formulations.
This capability is demonstrated with three multi-objective optimization tasks:
\begin{enumerate}
  \item \textit{Pareto Frontier Optimization.}- An LLM aims to identify solutions that are non-dominated across the considered objectives, effectively balancing trade-offs between the targets (for example, HOMO-LUMO gap and polarisability). 
  A sentence, "My design objective is to expand the Pareto frontier (maximizing) of my TMCs spanned by two properties, HOMO-LUMO gap and polarisability", is used in the prompt to achieve this design objective.
  \item \textit{Maximizing both HOMO-LUMO Gap and polarisability.}- An LLM aims to propose TMCs that increase both the HOMO-LUMO gap and the polarisability. 
  A sentence, "My design objective is to find TMCs that simultaneously maximize both HOMO-LUMO gap (> 4 eV) and polarisability (> 400 au)", is added in the prompt.
  \item \textit{Maximizing polarisability at a small HOMO-LUMO gap.}- An LLM aims to propose TMCs with extremely large polarisability that has a small HOMO-LUMO gap (< 1 eV). 
  A sentence, "My design objective is to design TMCs with maximized polarisability (> 450 au) and minimized HOMO-LUMO gap (< 1.0 eV)", is included in the prompt.
\end{enumerate}

\textit{Generating new ligands and TMCs. }
Besides asking LLMs to propose TMCs with optimized properties in a confined space (for example, the 1.37M TMC design space), one would expect unlocking the generative power of LLMs by requesting LLMs to generate new ligands and TMCs without constraining them to only use a small pool of ligands.
Provided the original 50 ligands and the top-100 TMCs made by these ligands in Pd(II) square planar geometry, two different setups are investigated in this work:
\begin{enumerate}
  \item \textit{Generating a new ligand pool in one shot.}- An LLM is asked to generate 10 new ligands: five neutral and five monoanionic. The key criterion was to ensure these new ligands were distinct from those in the existing pool of 50 ligands. Additionally, these ligands were designed to further optimize specific objectives (see Fig. \ref{fig:gen_tmc}a). To achieve this, the prompt provided to the LLM included the original 50 ligands' molecular structures in SMILES format, along with relevant data such as charges and connecting atom information. Furthermore, examples of TMCs with their calculated properties were included to guide the LLM's generation process. The new chemical space was constructed by enumerating all possible combinations of these 10 newly generated ligands, applying the same constraint on the total charge (–1, 0, or 1) as the existing pool. This process led to the creation of approximately 2,200 new TMCs. Finally, these new TMCs were evaluated using the same procedure as for the original 1.37 million TMC space, ensuring consistency in evaluation methodology (refer to \textit{\nameref{method:EvalTMC}}).
  \item \textit{Generating new TMCs with new ligands in an iterative manner.}- An LLM is asked to generate 10 new TMCs, potentially using new ligands that are outside of the 50-ligand pool, that would further optimize certain objectives (Fig. \ref{fig:gen_tmc}b).
  These 10 new TMCs are decomposed to obtain a certain number of new ligands.
  The new ligands, together with the new TMCs, are used to update the known information in the prompt.
  Then a new iteration follows.
  This evolutionary optimization process is repeated for a total of 20 iterations for single property optimization tasks (that is, maximizing polarisability) and 40 iterations for multi-objective optimization tasks (that is, maximizing both HOMO-LUMO gap and polarisability).
\end{enumerate}

\paragraph{Prompt Engineering.}\label{method:PE}
Writing good prompts is essential to guide LLMs to behave properly and exploit their performance.
We find, however, the performance of LLM-EO is not very sensitive to the prompt, as long as the description of known information and task is clear.
In practice, we tune the prompt against claude-3.5-sonnet to remove edge cases and confirm the output is in a proper format.
The prompt is then directly used to interact with o1-preview without any changes.

\paragraph{1.37M square planar TMC space.}\label{method:1.37MSpace}
With the initial 50-ligand pool introduced by Kneiding et al.\cite{Kneiding2024Apr}, a space of 1.37M Pd(II) square planer TMCs is constructed considering the total charge being -1, 0, or 1.
The acyclic symmetry of square planer TMCs, meaning Pd(L$_1$)(L$_2$)(L$_3$)(L$_4$) is the same as Pd(L$_2$)(L$_3$)(L$_4$)(L$_1$), Pd(L$_3$)(L$_4$)(L$_1$)(L$_3$), and Pd(L$_4$)(L$_1$)(L$_2$)(L$_3$), is considered when building this 1.37 TMC space.
It is worth noting that the 1.37M TMC dataset from Kneiding et al. was released on Feb 15, 2024, which is beyond the knowledge cutoff of OpenAI (Oct 2023) and close to the cutoff of claude-3.5-sonnet 
 from Anthropic (Apr 2024), minimizing the risk of direct data leakage.

\paragraph{Evaluating new TMCs.}\label{method:EvalTMC}
Similar to Kneiding et al.\cite{Kneiding2024Apr}, a new Pd(II) square planar complexes is generated with molSimplify by specifying the metal center, coordinating ligands represented by SMILES string, and their connecting atom indexes.
GFN2-xTB is then used to optimize the geometry generated by molSimplify and calculate HOMO-LUMO gap and polarisability.
TMCs that result in bad geometry either due to atom clashing or change of connectivity during the geometry optimization are considered as failed attempts, which are included in the prompt with reasons of failure to guide LLMs learning from failure\cite{LearningFromFailure}.

\section*{Code and data availability}
Code and data are currently under review and will be available as a open source repository on github.

\section*{Author contributions}
J.L. and Z.S.: methodology, software, validation, investigation, data curation, writing of original draft, review and editing, and visualization. 
Q.Z.: dataset, data curation, and review and editing.
Y.D.: methodology, dataset, and review and editing.
Y.C.: software and review and editing.
H.J.: review and editing and funding. 
C.D.: conceptualization, methodology, software, validation, investigation, data curation, writing of original draft, review and editing, visualization, and funding.

\section*{Acknowledgement}
We would like to thank our entire team from Deep Principle for helpful discussions and support.
C.D. thanks Haorui Wang, Jingru Gan, Yanjiao Zhu, and Peichen Zhong for valuable discussions.

\section*{Competing interests}
J.L., Z.S., H.J., and C.D. are co-inventors on a provisional patent application that incorporates discoveries described in this manuscript.

\bibliographystyle{naturemag_doi}
\bibliography{main.bib}

\clearpage

\setstretch{1}

\end{document}